# Photometric and Hα Studies of Two Extreme Mass-Ratio Short Period Contact Binaries in the Direction of Open Cluster Praesepe

**J. Rukmini and D. Shanti Priya**

Department of Astronomy, Osmania University, Hyderabad, Telangana-500007.

**Received:** 07 April 2016; **Revised:** 19 April 2016; **Accepted:** 00 April 2016

**Abstract:** We present the high precession photometric studies in V band and spectroscopic studies centered around Hα line for two extreme mass-ratio short period contact binaries ASAS J082243+1927.0 (V1) and ASAS J085710+1856.8 (V2). The variables in study are in the direction of open cluster Praesepe. From the light curve analysis V1 was found to be a low mass-ratio over-contact binary of A-type, with a mass ratio q~0.121 and fill-out factor f~72%, and V2 was found to be a low mass contact binary of W-type, with a mass-ratio q~1.29 and fill-out factor f~11% (marginal contact). The study of Hα absorption line profile of both the variables shows variation in equivalent widths (EWs) with orbital phases. The mean EWs of the Hα line were obtained as 1.6±0.13 Å and 1.18±0.12 Å for V1 and V2 respectively. The variation of Hα absorption with respect to phase is explained to be due to chromospheric activity in V1, as evident from the O'Connell effect and that due to chromospheric flares/winds or photospheric source in V2. The parameters for both the binaries are also studied with respect to the large sample of well-studied contact binaries and possible mechanisms for merger in contact binaries of both low and high mass ratios emphasized.

## INTRODUCTION

Contact binaries (also known as W UMa variables) are a unique class of eclipsing binaries with typical period of 0.2-1.0 d, consisting of mostly main sequence components with spectral types ranging from A-K type. They share a common convective envelope resulting in geometrical effects as evident from their light curves.

Contact binaries (CBs) play an important role in the study of structure and evolution of stars. They provide reliable basic parameters when studied through various photometric and spectroscopic techniques. The photometric investigations of CBs have led to reliable characterization of the components in terms of their masses, radii, effective temperatures, luminosities, chromospheric activities and also giving an insight into probable merger processes through period analysis (Liu et al. 2016, Qian et al. 2015, Y.C. Joshi et al. 2015, Kandulapati et al. 2015, Shanti et al. 2013, 2014). In addition, the spectroscopic investigations of CBs focused on Hα line profile (which is observed to be a primary indicator of chromospheric activity) not only provide substantial information of magnetic associated activity in one or both the components but also derive their variation with respect to orbital phases (Kandulapati et al. 2015, Diana & Dragomir 2011, Vilhu & Maceroni 2007).

About 100 eclipsing binaries were categorized by Pepper et al. (2008), using light curves obtained for 66,638 stars in NGC 2632 (Praesepe). The current work deals with the photometric and spectroscopic study done on two of the short period variables, which were studied for the first time by our group (Shanti et al., 2013; Kandulapati et al., 2015). Due to their small orbital periods and shape of the light curves, they were classified as W UMa type of binaries. The two variables have shown extreme photometric properties; hence we have chosen them for further study to understand the complimentary characteristics, which may in different scenario lead to merger in binaries. The details of the observational data acquired and the method of analysis used for reduction and analysis of this data are explained in section 2. The parametric results obtained through the photometric analysis are discussed in section 3, while that from spectroscopic analysis in section 4. The discussion on the results obtained and the underlying mechanism of probable mergers along with the comparative statistical study with respect to the collection of well-studied contact binaries is presented in section 5.

## DATA COLLECTION AND ANALYSIS

The V band photometric observations were made for both the variables V1 and V2 with the 2.0 m telescope of the IUCAA - Girawali Observatory (IGO), India. The observations for variable V1 were performed during three nights, on January 19, 20 and 22, 2013 and for variable V2, during two nights, on February 26 and 27, 2012, using IUCAA Faint Object Spectrograph and Camera (IFOSC). A total of 418 and 289 frames in the V band were obtained, for variables V1 and V2 respectively. Reduction of photometric data was performed using aperture photometry with the *apphot* package available in IRAF software. The period and times of minima of the variables were determined using Peiord04 package (Lenz and Breger, 2005) and the Kwee and Van Woerdens method (1956) respectively.

Spectroscopic observations for both the variables V1 and V2 were carried out using Vainu Bappu Telescope (VBT) at Vainu Bappu Observatory (VBO) on December 5 and 6, 2013. The telescope was equipped with an Opto-Mechanics Research (OMR) spectrograph along with a detector of 1K × 1K CCD.

The spectra obtained subtends a range of 3000A° window centered around Hα line (6563 Å) with a dispersion of 600 lines/mm grating, resulting in a resolution of 2.6A°/pixel. The exposure time was 35–40 min for both the variables respectively along with the spectrophotometric standard (BD+08 2015). Fe-Ne comparison spectra was observed for wavelength calibration. A total of four spectra were obtained at various phases for each of the variable. The standard available package ONEDSPEC in IRAF was used to reduce the spectroscopic data. The spectra obtained after reduction were then calibrated and normalized for further studies using standard procedures. From the obtained spectra, the equivalent widths of Hα profile were computed.

The intrinsic equivalent widths were calculated using the following relation.

$EW_p^i = a \times EW_p$ & $EW_s^i = a \times EW_s$, where $a = 1 + q^{0.92} (T_s/T_p)^4$ (Vilhu & Maceroni 2007)

$EW_p$, $EW_s$ are measured equivalent widths and $EW_p^i$, $EW_s^i$ are intrinsic equivalent widths.

## PHOTOMETRIC RESULTS

While various space missions are looking for habitable extra-solar planets, photometric studies of short period contact binaries provide constraints on the theories of formation and evolution of low mass stars and that of the associated planets. And one of the hypothesis explaining the origin of hot jupiters in extra-solar systems is that they are formed in protoplanetary disks by the merger of short period, low mass binaries via “magnetic braking” (Martin et al., 2011).

The ephemeris determined for the binaries from the reduced photometric data are given by the following expressions. The numbers in parentheses are the errors in terms of the last quoted digits.

For variable V1,

Min I = 2456312.2997(89) + $0^d$.28000(2) E

For variable V2,

Min I = 2455984.3552(12) + $0^d$.29100(3) E

The ephemeris determined and the periods obtained for both the binaries are same as the one reported earlier (Pepper et al., 2008). The photometric analysis was done by using the W-D program with an option of non-linear limb darkening via a square root law along with many other features (van Hamme &Wilson 2003) and the best solutions were used to obtain basic parameters of the variables as listed in the **Table 1**. Both the variables V1 and V2 show typically short periods and have a visual magnitude of about 10-11. They are found to be in the direction of cluster Presepae but are not likely members of the cluster (Pepper et al. 2008). However, they show quite different properties, which indicates that they have a different age and evolutionary status. From their colors, the spectral types have been found to be as G0 for V1 and G6-K0 for the variable V2 (based on Allen’s tables, Allen, 2000).

The solution converged for V1 at a mass-ratio q ~ 0.12, inclination ~ 76° and a fill-out factor of about f ~ 72%, which implies that, it has high geometrical contact. The parameters clearly suggest that it is a deep low mass ratio over contact binary (DLMR). The DLMR contact binaries typically show $q \leq 0.25$ and $f \geq 50\%$ and are theorized to be possible progenitors of FK Com type stars and Blue Stragglers (Qian et al., 2006). The solution converged for V2 at a mass-ratio of q ~ 1.29, inclination ~ 76$^o$ and a fill-out factor of about f ~ 11% only suggesting that it is a contact binary with shallow geometrical contact configuration.

**Table 1.** The photometric parameters derived for the variables V1 and V2 using Wilson-Devinney (WD) method

| Fundamental Parameter | Variable V1 | Variable V2 |
|---|---|---|
| Period (P) | 0.28 d | 0.291 d |
| Spectral type | G0 | G6-K0 |
| Primary temperature T1 | 5960 K | 5477 K |
| Secondary temperature T2 | 6038 K | 5309 K |
| Mass Ratio ($q = m_2/m_1$) | 0.106 | 1.29 |
| Inclination (i) | 76.58 | 76.22 |
| Fill-out factor (f) | 0.72 | 0.11 |
| Class | A-type | W-type |
| Spot activity (O'Connell effect) | Present | Absent |

The components of both the variables show a low temperature difference, indicating that they are in good thermal contact. Based on the best fit solutions, the LC program of WD code was run to get the theoretical light curves. The observed and the theoretical light curves are shown in the **Figure 1.**

The fit in the light curves is found to be satisfactory. Though the derived values of the inclination are same for both the variables, the light curve of V1 distinctly shows a totality at secondary minima and asymmetry in the maxima (O'Connell effect), which as derived from the solution, is due to their relative absolute dimensions when compared to their orbital geometry. Many contact binaries show asymmetry in their light curves around maxima, i.e. around phases 0.25 and 0.75 (O'Connell, 1951) which is generally explained as due to presence of cool/hot spots on one of the components of the binary (Kalomeni et al., 2007).

From the solutions, the asymmetry in V1 is defined to be due to the chromospheric activity like presence of a cool spot on one of the components, which could be migrating. Hence, a varying chromospheric activity is suggested for V1. However, no spot activity could be concluded from the light curve of V2. The geometrical configurations of the variables are shown in **Figure 2**. The fundamental parameters derived from the solutions suggest that V1 is A-subtype whereas V2 is W-subtype W UMa binary and the components of V2 are considerably undersized and underluminous.

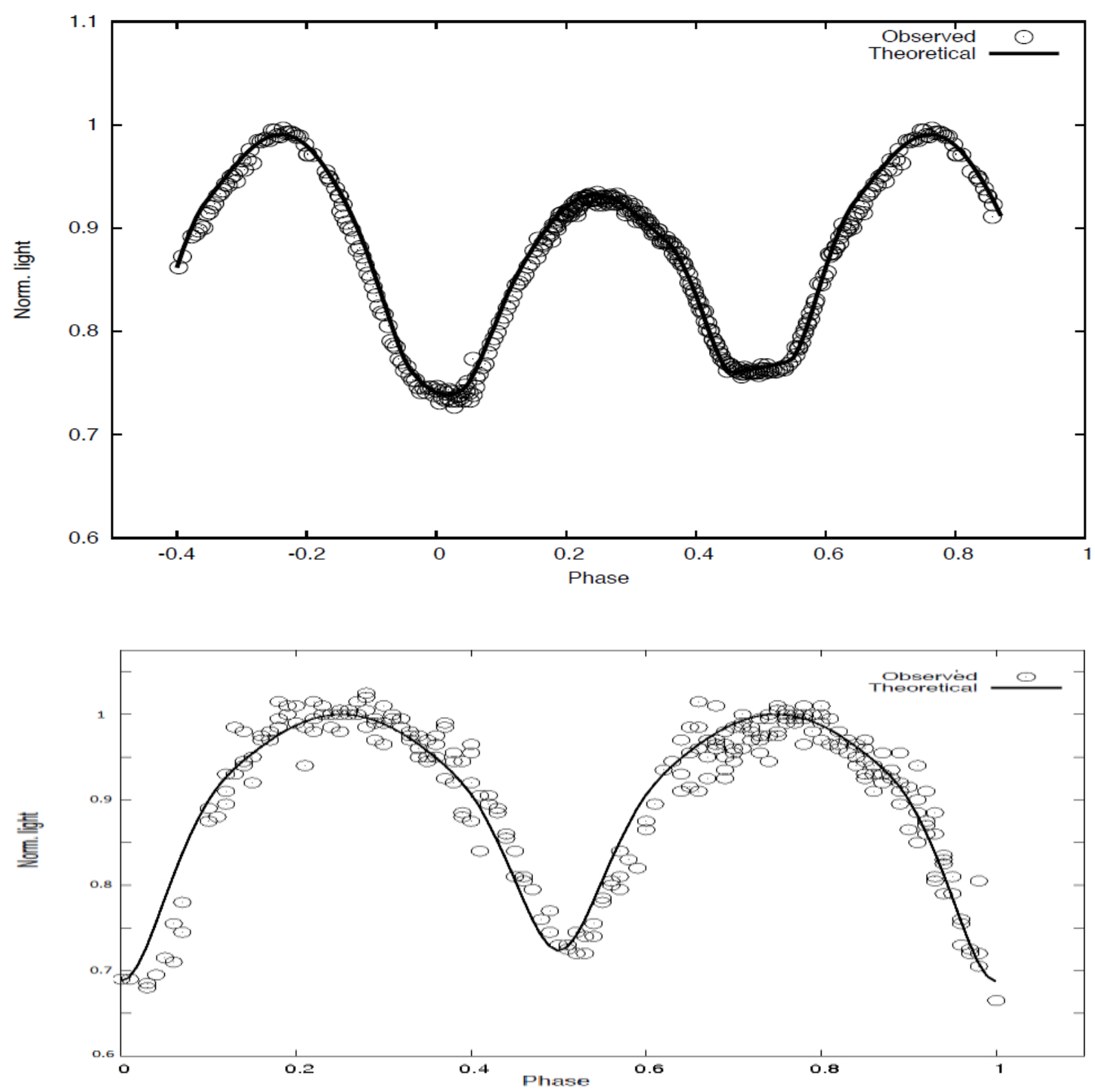

**Figure 1.** Observed and Theoretical light curves of the variable V1 (top panel) and V2 (bottom Panel). V1 shows a totality at phase 0.5.

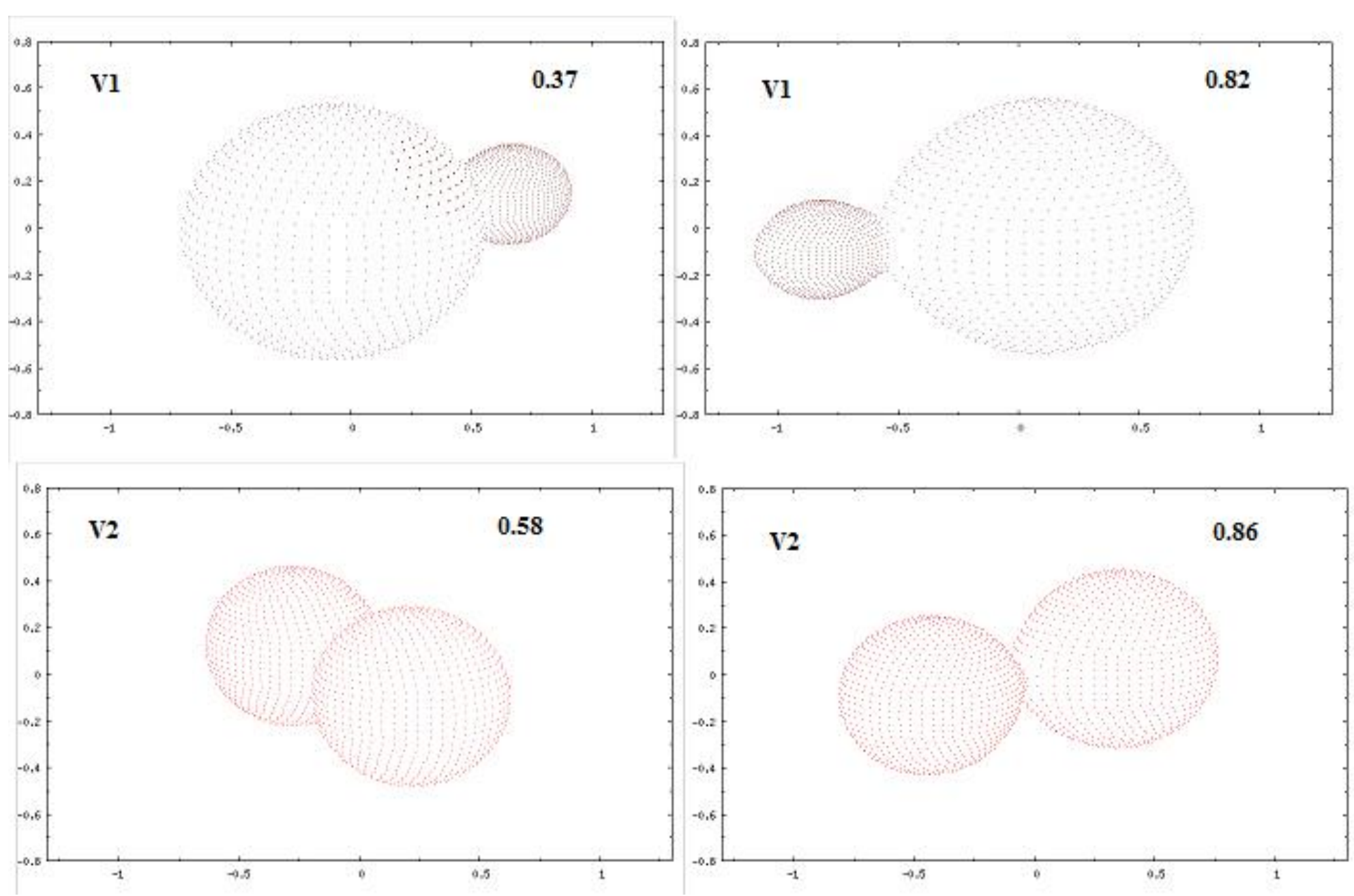

**Figure 2.** The geometrical configuration of both the variables V1 (top panel) and V2 (bottom panel) at two different phases. The cool spot can be seen for V1 at phase 0.37.

## SPECTROSCOPIC RESULTS

Study of Hα line profile is a valuable tool for understanding activity in stars, which is as well followed for binaries. The first detailed spectroscopic study of Hα line was performed by Barden (1985) on four W UMa binaries proving the Hα line to be a strong signature of the magnetic-associated activity in these systems.

The Hα studies of two contact binaries done by Vilhu & Maceroni (2007) established that the average equivalent widths of Hα lines were close to the saturation border and the primary components show excess Hα emission indicating enhanced chromospheric activity. The radial velocity solutions were obtained by Diana & Dragomir (2011), for FI Boo on the basis of Hα observation in 2007-08. They concluded that the deeper Hα absorption in 2007 could be due to the presence of a transient compact source, as the observed absorption was not correlating with the orbital phases.

While the Hα line profile for the variable V1 was studied at phases 0.15, 0.37, 0.50 and 0.82, it was studied at phases 0.27, 0.58, 0.68 and 0.86 for the variable V2. **Figure 3** shows the variation of Hα line at various phases for the variables along with the standard. The equivalent widths of the Hα line observed at the four phases for both the variable are as listed in **Table 2**.

The mean equivalent widths (EW) for V1 and V2 were found to be 1.60 ± 0.13 A and 1.18 ± 0.12 A respectively. The derived mean EWs for both the variables are in good agreement with the value derived by Cram & Mullan (1985) for stellar models with corresponding effective temperatures.

**Table 2.** The equivalent widths (EW) of Hα line at observed phases for the variables V1 and V2.

| Variable V1 | | Variable V2 | |
|---|---|---|---|
| Phase | EW | Phase | EW |
| 0.15 | 1.246 | 0.27 | 1.090 |
| 0.37 | 2.428 | 0.58 | 0.643 |
| 0.50 | 0.903 | 0.68 | 0.675 |
| 0.82 | 1.833 | 0.86 | 1.820 |

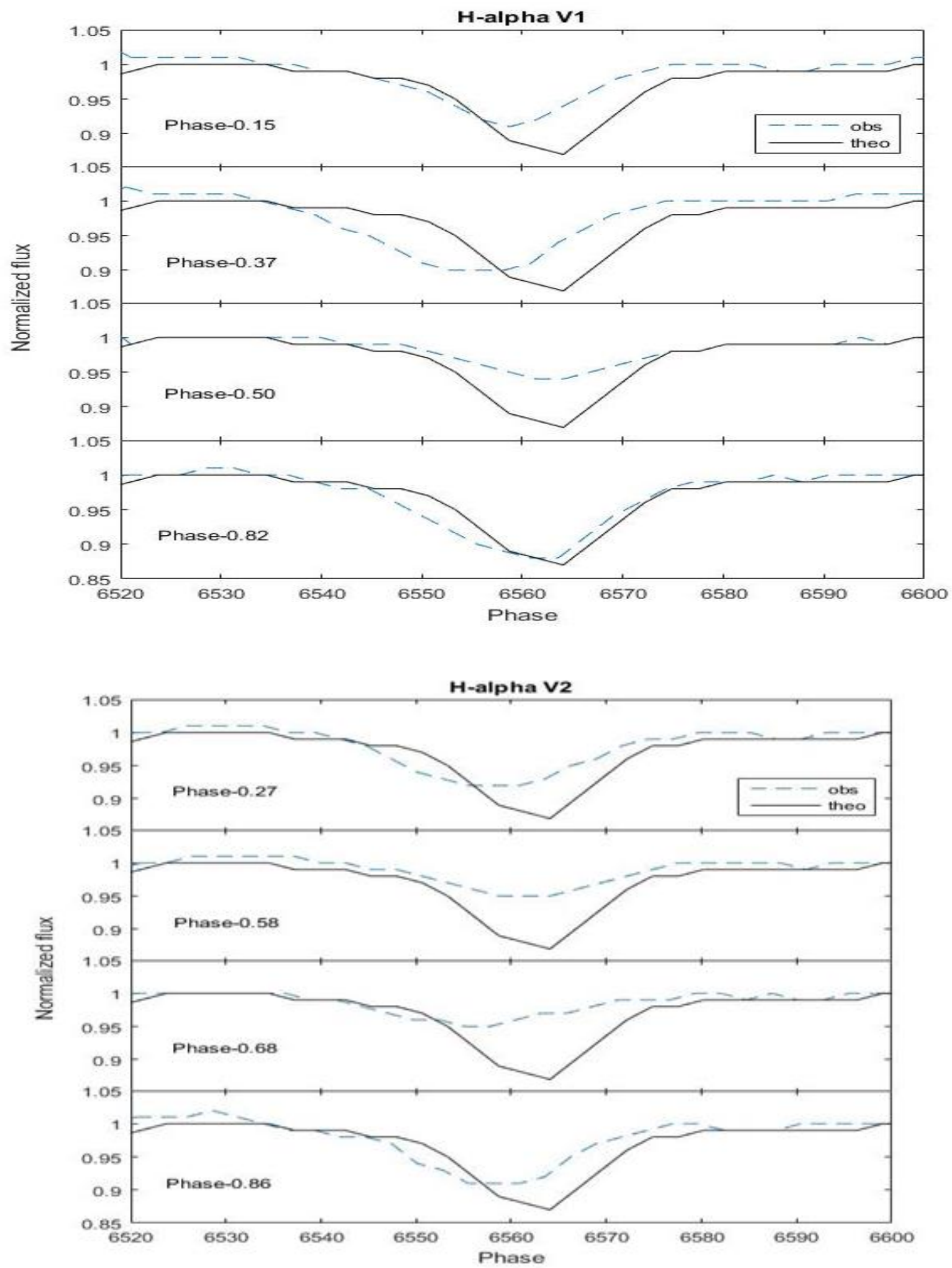

**Figure 3.** Variation of Hα line profile at various phases for the variables V1 (top panel) and V2 (bottom panel) along with the standard star.

## DISCUSSION

As the W UMa type systems outnumber all other categories of eclipsing binaries, their study becomes important in understanding binary formation and evolution theories (Yakut & Eggleton, 2005), the merger (Tylenda et al., 2011) and stellar dynamo process (Qian et al., 2005), influence on the galactic structure (Rucinski, 2002), their use as distance indicators (Rucinski & Duerbeck, 1997). Incidentally, the study of Hα line in contact binaries, in addition to photometric studies has proved to provide a strong evidence for the presence of chromospheric activity (ex: spots, flares etc).

The photometric solution of the variable V1 suggests that it is a deep low mass ratio overcontact binary, a characteristic of A-subtype W UMa binary. But due to the relatively higher temperature of the secondary, it is classified as W-subtype W UMa binary. The present study when compared to the previous studies shows that, the system may be having a migration of spots on its surface i.e., variable O'Connell effect (a strong signature of chromospheric activity). The same reasons may justify the transition from A-subtype to W-subtype and vice-versa, as observed in some other contact binaries FG Hya, V802 Aql, V902 Sgr (Qian & Yang 2005, Zola et al. 2010, Samec et al. 2004, Yang et al. 2008, Samec & Corbin 2002). In the spectroscopic studies of variable V1, it is observed that equivalent width of the Hα absorption line is varying with phase as observed in table 2. The Hα line is weakly filled in at all phases except near secondary eclipse. That is, it shows a possibly filled-in effect at phase 0.5, indicating the presence of chromospheric activity and the primary component to be the source. There also exists a correlation between the orbital period and the Hα line equivalent width of the primary component of contact binaries.

The photometric solution of the variable V2 clearly suggests that it is a low mass shallow/marginal contact binary of W-subtype. And the light curve of the variable does not display any O'Connell effect. However, it is observed that the absorption profile is weakly filled-in at phase 0.62, mostly by chromospheric activity, while the Hα line shows a stronger filled-in absorption profile at phase 0.18. As seen from table 2, the EW of Hα line shows variation with phase, i.e. it's found to be minimum just before the egress of primary eclipse, increasing with the ingress of secondary eclipse and reaching maximum just before the egress of the secondary eclipse, suggesting the source of Hα absorption line to be on or near the primary component. It is not possible to substantially separate the contribution of each component of the binary from the limited observations of observed phases. Generally, one of the reasons for the variation or asymmetry in Hα profile could be chromospheric spots, winds, flares etc (Vilhu et al., 1991, Jetsu et al., 1993). But the absence of the O'Connell effect suggests that the variation in the EW of the Hα profile could also be due to photospheric absorption (Cram & Mullan 1985).

Thus by comparing the characteristics of both the variables, we observe them to be two extreme binaries. Both low mass and massive contact binaries are the most interesting objects for study as they have high probability for undergoing the process of mergers. Initially both the components of a close binary evolving into contact system are well inside their roche lobes. When the orbital angular momentum (AM) decreases, one or both the components overflow their roche lobes, resulting into a contact configuration with a common envelope around the components. Further DLMR over contact binaries are supposed to be the progenitors of FK Com-type and blue straggler stars. The gradual decrease of mass ratio makes them evolve into single fast rotating stars due to the induced Darwin's instability when the orbital angular momentum is more than three times the spin angular momentum (Hut 1980). The high degree of over contact may also cause a dynamical instability, leading to mergers (Rasio & Shapiro 1995). The mass transfer from primary to secondary component of a binary system results into short period high mass ratio systems. Moreover, as period decreases due to mass & AM transfer between the components at lagrangian point L1 and their loss from the system at L2, the critical Roche Lobes shrink causing 'f' to increase and finally the variable evolves as a single rapid rotator. Hence high 'f' values can be associated with low mass ratio systems. We have plotted the period-colour relation for 110 over contact binary systems from the data given by Terrell et al. (2012). The line in **Figure 4** shows the short period blue envelope (SPBE) theoretical line i.e. $(B\text{-}V)_o = 0.04 \times P^{-2.25}$ (Rucinski 1997) and the variable V1 is found to be above it. Any over contact binary above SPBE line has to further evolve towards longer period and cooler temperature (towards lower right in Figure). The position of the variable V1 is suggestive that it is an important over contact binary due to its low mass ratio and high fill-out factor and could be at a stage of merger, potentially forming into a FK-Com/Blue Straggler type star.

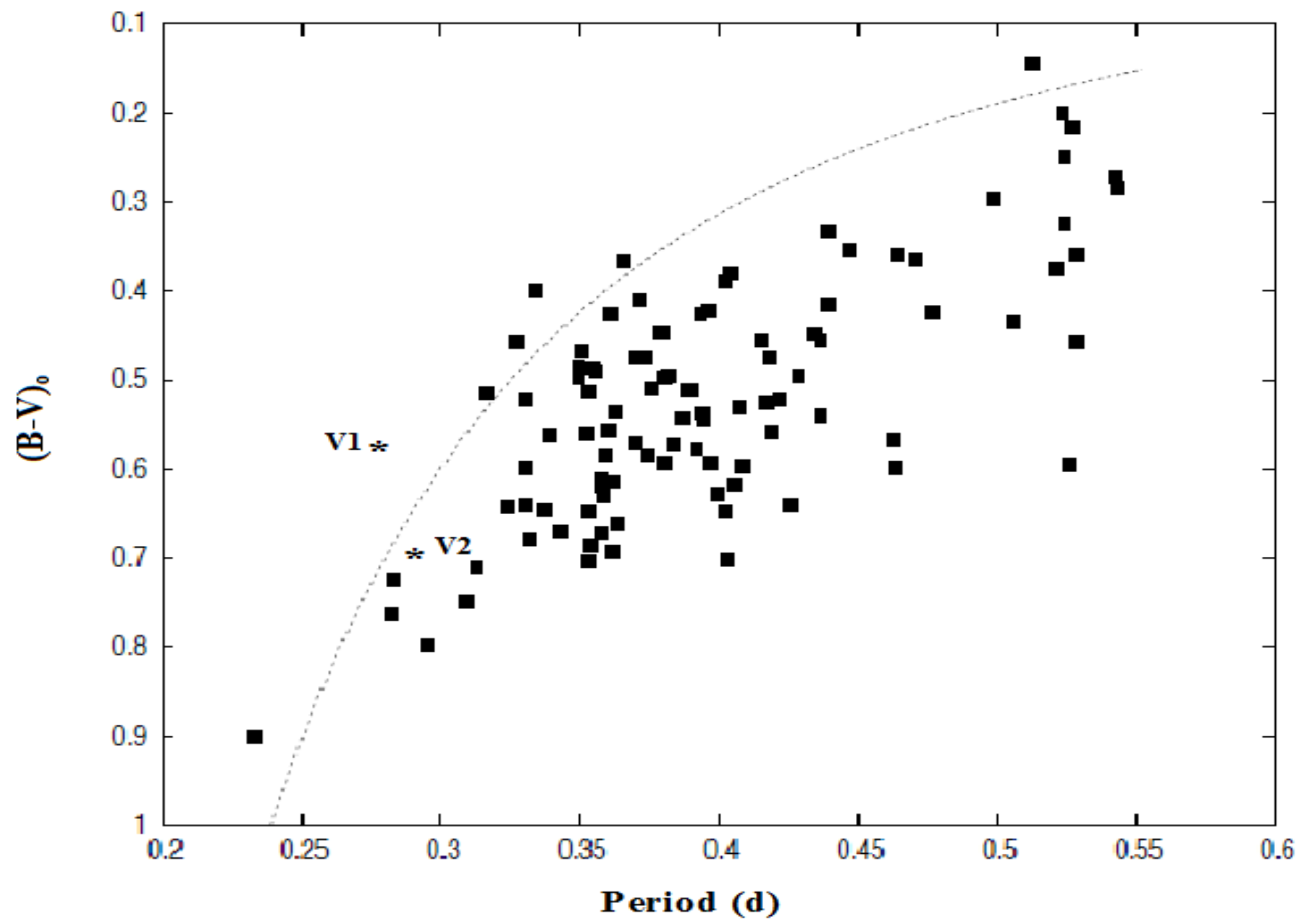

**Figure 4:** The period-colour relation for 110 over contact binary systems from the data given by Terrell et al. (2012). The dotted line represents the short period blue envelope (SPBE) theoretical line. The asterisk marks represent the variables in study.

The position of variable V2 is also shown on the plot, however the above explanation cannot be extended to it as it does not possess low mass-ratio and high fill-factor. Hence we tried to understand its evolutionary state in a different scenario. Low Mass Contact Binaries (LMCBs) with later spectral types show highest level of chromospheric/coronal activity (Gazeas & Stepien 2008). They lose AM and mass by magnetized wind similar to that of single active stars. From the model given by Gazeas & Stepien (2008) the orbital AM for the variables were calculated using the relation

$$H_{orb} = (1.24 \times 10^{52})\, M^{5/3} P^{1/3} q\, (1 + q)^{-2}$$

Where P is period (in days), $H_{orb}$ is orbital AM (in cgs units), $M = M_1 + M_2$ is the total mass of the system (in solar units) and q is mass ratio. The $H_{orb}$ value obtained for variables V1 and V2 are $4.294 \times 10^{51}$ and $1.455 \times 10^{51}$ respectively. **Figure 5** shows the plot of observed periods with orbital AM for LMCBs (Gazeas & Stepien 2008) along with other W UMa type systems (Gazeas & Stepien, 2008) and the values for V1 and V2 are overplotted. In this plot, the position of V2 clearly indicates that it is in the upper marginal region for LMCBs. As per the models discussed by Gazeas & Stepien (2008), the variable could be a progenitor for a merger even with a higher mass ratio, because in LMCBs roche lobe expansion across main sequence is slow and AM loss dominates causing coalescence of the binary even before the mass ratio becomes very low. The variability of EW of Hα profile with phase, which is suggestive of surface activities other than spots can explain the AM and mass loss from the variable V2, indicating it also to be a potential candidate for mergers along with that of variable V1.

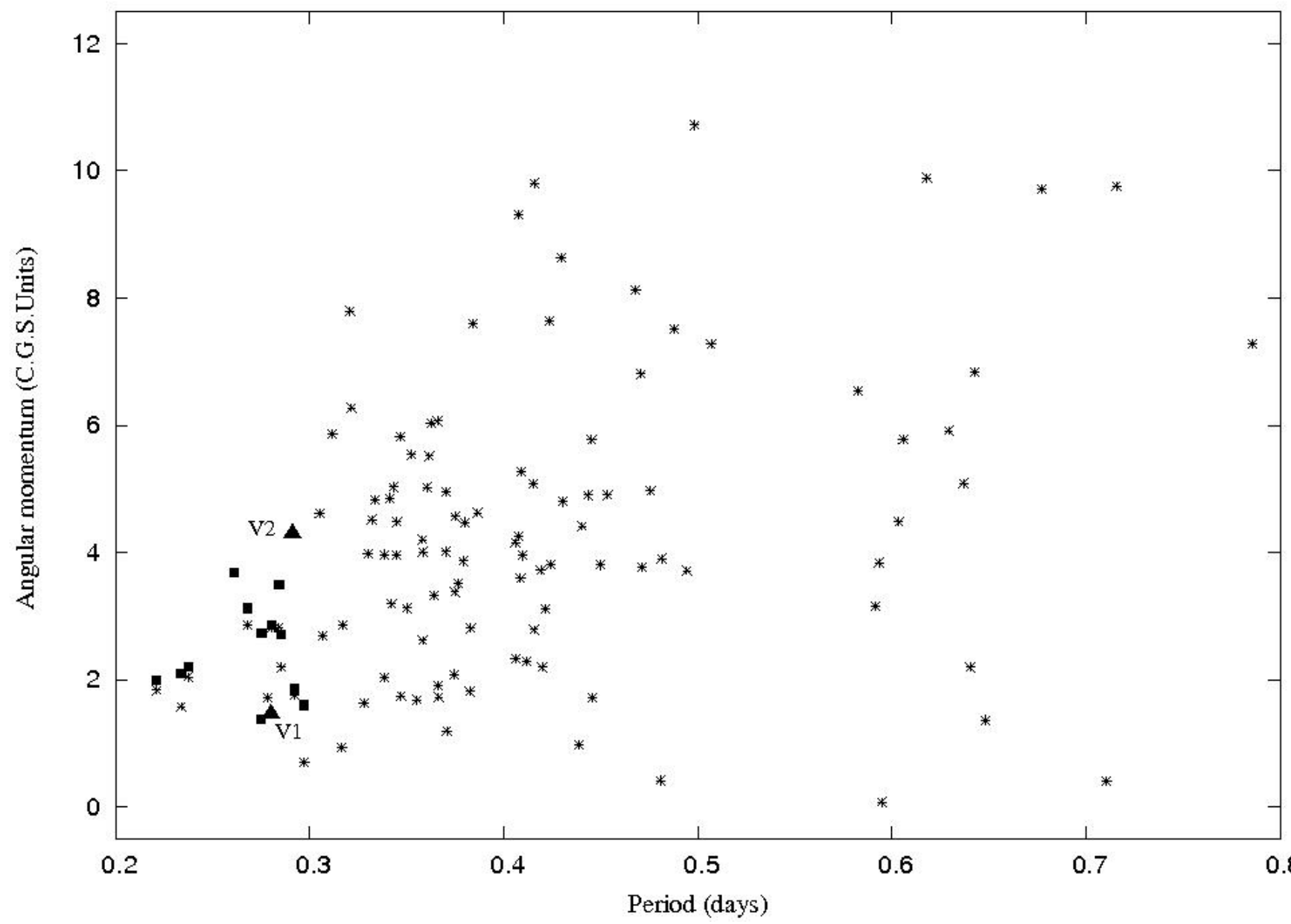

**Figure 5:** Observed periods with orbital AM for LMCBs (Gazeas & Stepien 2008) shown as filled squares along with other W UMa type systems (Gazeas & Stepien, 2008) shown as asterisk and the values for V1 and V2 over plotted, shown as triangles.

## CONCLUSION

Both the photometric and spectroscopic observations done in the present study were limited by time and hence long term observations can help in deriving further details of the two variables, like the ambiguity of rate at which the mass and AM loss is taking place in V1, which can be concluded from period variation studies and the nature of the source producing Hα line in V2, where no O'Connell effect is observed.

Studies of contact binaries are one of the best ways to derive the absolute parameters of stars of different spectral classes, luminosity classes and evolutionary stages, including for those, which can host habitable planets. Due to their close proximity, mutual gravitational interaction indeed influences not just the matter but also the radiation coming from both the components. Multiple observing projects like having simultaneous photometric and spectroscopic observations at multi-wavelength bands can infact reveal additional information of this category of eclipsing binaries, which have a high incidence (number densities) in our and other galaxies.

**Corresponding author: J. Rukmini**

Department of Astronomy, Osmania University, Hyderabad, Telangana-500007.